\documentclass[aps,prd, 9pt, onecolumn,nofootinbib,showpacs]{revtex4-1}
\usepackage{subeqn}
\usepackage[usenames,dvipsnames]{xcolor} 
\usepackage[normalem]{ulem}
\usepackage{graphicx}

\begin{document}

\title{Overcharging dilaton black holes in 2+1 dimensions to extremality and beyond}
\author{Koray D\"{u}zta\c{s}}
\email{koray.duztas@ozyegin.edu.tr}
\affiliation{Department of Natural and Mathematical Sciences,
\"{O}zye\u{g}in University, 34794 \.{I}stanbul, Turkey}
\author{Mubasher Jamil}
\email{mjamil@zjut.edu.cn (corresponding author)}
\affiliation{Institute of Theoretical Physics and Cosmology, Zhejiang University of Technology, Hangzhou 310023, China}
\affiliation{Department of Mathematics, School of Natural Sciences (SNS), National
University of Sciences and Technology (NUST), H-12, Islamabad 44000, Pakistan}
\affiliation{Canadian Quantum Research Center, 204-3002 32 Ave, Vernon, BC V1T 2L7, Canada}

\begin{abstract}
We test whether static charged dilaton black holes in 2+1 dimensions can be turned into naked singularities, by sending in test particles from infinity. We derive that  overcharging is possible and generic for both  extremal and nearly extremal black holes. Our analysis also imply that nearly extremal charged dilaton black holes can be continuously  driven to extremality and beyond, unlike nearly extremal Ba\u{n}ados-Teitelboim-Zanelli, Kerr and Reissner-Nordstr\"{o}m black holes which are overspun or overcharged by a discrete jump. Thus the weak form of the cosmic censorship conjecture and the third law of black hole thermodynamics are both violated in the interaction of charged dilaton black holes in 2+1 dimensions, with test particles. We also derive that there exists no points where the heat capacity vanishes or diverges in the transition from black holes to naked singularities. The phase transitions that could potentially prevent the formation of naked singularities do not occur. 
\end{abstract}
\maketitle
\newpage
\section{Introduction}
Following the discovery of the Ba\u{n}ados-Teitelboim-Zanelli (BTZ) metric~\cite{btzmetric}, the black hole solutions in $(2+1)$ dimensions have intrigued physicists over the last three decades both in the context of general relativity and low energy string theory. Motivated by the considerable simplification of the field equations, classical, quantum and thermodynamic  properties of the lower dimensional black holes have been extensively studied. The BTZ black hole is also a solution to low energy string theory. The Einstein-Maxwell dilaton action with different values of couplings can yield different black hole solutions in 2+1 dimensions.
In this respect Chan and Mann considered the Einstein-Maxwell dilaton action with arbitrary couplings $a$, $b$ and $B$\cite{mann}:
\begin{equation}
S=\int d^3x \sqrt{-g}\Big[ R-\frac{B}{2}(\nabla\phi)^2-e^{-4 a\phi}F_{\mu\nu}F^{\mu\nu}+2e^{b\phi}\Lambda\Big],
\label{action1}
\end{equation}
where $\phi$ represents the dilaton field, and $\Lambda$ is the cosmological constant while we set $B=8$ \cite{mann}. If we assume $\phi=k \ln(r/\beta)$, the couplings are fixed by the rule $4ak=bk=N-2$ so that the action (\ref{action1}) is conformally related to the low energy string action in 2+1 dimensions \cite{mann}. The field equations yield exact black hole solutions with mass $M$, electric charge $Q$ in the form
\begin{eqnarray}
 ds^2=&&-\left(-2Mr^{(2/N)-1}+ \frac{8\Lambda r^2}{(3N-2)N} + \frac{8Q^2}{(2-N)N}\right)dt^2  \nonumber\\&&
+\frac{4r^2 dr^2}{N^2 \gamma^{(4/N)}\left(-2Mr^{(2/N)-1} + \frac{8\Lambda r^2}{(3N-2)N} + \frac{8Q^2}{(2-N)N}\right)} +r^2 d\theta^2 .
\label{metric0}
\end{eqnarray}
Mann and Chan has used the following notation for the electric charge of black hole: $Q^2=\frac{q^2\beta}{\gamma^2}$, here the arbitrary couplings can be set to unity as a special case. These solutions represent black holes surrounded by an event horizon provided that $2>N>(2/3)$~\cite{mann}.  
In this work we focus on the particular solution for $N=1$, $b=4a=4$, $\gamma=1$ and $k=-1/4$. The resulting metric describes a static charged dilaton black hole  in (2+1)-dimensions \cite{fer0}
\begin{eqnarray}
 ds^2&=&-(8\Lambda r^2 - 2Mr  +8Q^2)dt^2 +\frac{4r^2 dr^2}{( 8\Lambda r^2 -2Mr +8Q^2)}+ r^2 d\theta^2, \nonumber \\
&& \phi=\left (\frac{-1}{4}\right) \ln \left( \frac{r}{\beta} \right); \quad F_{01}=e^{4\phi}\frac{Q}{r} .\label{metric}
\end{eqnarray}
There exists a singularity at $r=0$ which is covered by an event horizon, provided that $M\geq 8Q\sqrt{\Lambda}$.  For later calculations, we assume $\beta=1$. The spatial coordinates of the inner outer event horizons is given by
\begin{equation}
r_{\pm}=\frac{M \pm \sqrt{M^2-64Q^2\Lambda}}{8 \Lambda}.
\end{equation}
The spacetime represents a black hole with Hawking temperature
\begin{equation}
T_{\rm{H}}=\frac{M}{4\pi r_+}\sqrt{1-\frac{64 Q^2 \Lambda}{M^2}} ,\label{hawt}
\end{equation}
which vanishes in the extremal case $M=8Q\sqrt{\Lambda}$. 

In literature, numerous aspects of lower dimensional charged dilaton black holes have been studied, including quasi normal modes~\cite{fer0}, Hawking radiation and other thermodynamical aspects \cite{iz,seb}. The black hole has also been used as a particle accelerator \cite{fer}. More recently, these black holes are explored in the framework of non-linear electrodynamics as well \cite{hendi}. We are interested to test the weak form of the cosmic censorship conjecture (wccc) for a charged 2+1 dimensional black hole. The conjecture was proposed by Penrose to maintain the deterministic nature of general relativity in the existence of a curvature singularity~\cite{ccc}. As the singularity theorems by Penrose and Hawking indicated that the formation of singularities is inevitable in gravitational collapse, Penrose conjectured these singularities to be hidden behind the event horizons of black holes. In this manner the observers at asymptotically flat infinity do not encounter any effects propagating out of the singularity, and the space-time maintains its smooth causal structure. The stability of the event horizon is also crucial to assign thermodynamical properties to a black hole. 

A general proof of wccc turned out to be elusive. Wald constructed a thought experiment where one perturbs  an extremal or a nearly extremal black hole with test particles or fields and checks if it is possible to destroy the event horizon ~\cite{wald74}. Though the notion of distant observers is not well defined in Anti-de Sitter (AdS) space-times, the stability of event horizons has been tested in (2+1) and higher dimensions by constructing Wald type thought experiments~\cite{vitor,rocha,gwak1,gwak2,btz,gwak3, J1,J2,dilat}. Another motivation for testing the validity of wccc in asymptotically AdS space-times arises from the AdS/CFT correspondence, which connects string theories formulated on AdS space-times to conformal field theories on the boundary. In particular if a BTZ black hole can be overspun beyond extremality, there exists states on the boundary theory rotating at speeds larger than the speed of light. In this respect Rocha and Cardoso constructed a thought experiment and claimed that BTZ black holes cannot be overspun and one should be distrustful of any process that would result in the overspinning of a black hole in AdS space-time~\cite{vitor}. However, it was later shown --by one of the authors-- that the process described by themselves can lead to overspinning  if one starts with a nearly extremal black hole~\cite{btz}. In this work we continue to probe the stability of event horizons  by testing the validity of wccc for charged dilaton black holes in (2+1) dimensions for $N=1$. We are also going to evaluate the validity of the third law of black hole thermodynamics, which states that the black holes cannot be continuously driven to extremality. 

\section{Charged dilaton black holes and wccc}

There exist  numerous Gedanken  experiments in literature where the validity of wccc is tested in the interactions of black holes with test particles~\cite{hu,jacob,gao,magne,higher,v1,dkn,overspin,emccc,duztas,toth,natario,hawk,mode,siahaan,taub-nut,w2,kerrsen,generic,gwak5,manko,chen1,chen2,string,J3}. In the test particle approximation, the geometrical structure of the spacetime is maintained, however small variations in the mass, charge and angular momentum parameters of the black hole are observed. These variations determine the final parameters of the spacetime, which could represent a black hole or a naked singularity. 

We assume that the test particles which are not absorbed by the black hole are scattered back to infinity without changing the parameters of the background geometry. There exists a lower limit for the energy of the test particle, which allows its absorption by the black hole. In addition, we point out that the dynamics of charged test particles in the vicinity of (2+1)-charged dilaton black hole is studied in \cite{fer} where it is demonstrated with the use of numerical properties of effective potential that the charged particle is absorbed by the black hole provided the energy of the particle is greater than a certain critical energy.

One can derive  the lower limit of energy by considering the dynamics of a test particle of mass $m$ charge $q$ in a curved background, described by the geodesic equations
\begin{equation}
\ddot{x}^{\mu}+ \Gamma^{\mu}_{~\rho \sigma}\dot{x}^{\rho} \dot{x}^{\sigma}=\frac{q}{m}F^{\mu \nu}\dot{x}_{\nu}.
\end{equation}
These equations follow from the Lagrangian
\begin{equation}
\mathcal{L}=\frac{1}{2}mg_{\mu \nu}\dot{x}^{\mu}\dot{x}^{\nu}+ qA_{\mu}\dot{x}^{\mu},
\end{equation}
where $A_\mu$ is the vector potential. For the metric (\ref{metric}), $F=dA$ implies that  ${\bf A}=(-Q/r) dt$. Since the given metric is static and endowed with a timelike Killing vector, the energy of the particle will be conserved
\begin{eqnarray}
E&=&-\frac{\partial \mathcal{L}}{\partial \dot{t}}=-mg_{00}\dot{t}-qA_0 \nonumber \\
&=&m\left(-2Mr+ 8\Lambda r^2 +8Q^2\right)\dot{t}+\frac{qQ}{r} .\label{E}
\end{eqnarray}
Therefore the energy of a test particle which crosses the event horizon at ($r=r_+$) should satisfy
\begin{equation}
E\geq E_{\rm{min}}= \frac{qQ}{r_+}. \label{emin}
\end{equation}
Eq. (\ref{emin}) determines the minimum energy  $E_{\rm{min}}$, which would allow a test particle to be absorbed by  the black hole.  Conventionally we assume that the spacetime is unperturbed and precisely described by the metric (\ref{metric}) before the initial time $t_{\rm{in}}$. At $t=t_{\rm{in}}$ we send in a test particle from infinity with energy $E>E_{\rm{min}}$ and charge $q \ll Q$. After the test particle is absorbed sufficient time has passed, we assume that the spacetime eventually settles down to a new solution with final parameters $M_{\rm{fin}}=M+E$ and $Q_{\rm{fin}}=Q+q$. Finally we check whether it could be possible for the final parameters of the spacetime to represent a naked singularity.

\subsection{Overcharging extremal black holes}

In this section, we start with an extremal black hole satisfying 
\begin{equation}
\delta_{\rm{in}}=M^2-64Q^2\Lambda=0. \label{deltainex}
\end{equation}
Notice that the spatial coordinate of the event horizon for the extremal black hole is
\begin{equation}
r_+ =\frac{M}{8\Lambda}=\frac{Q}{\sqrt{\Lambda}}. \label{rplus}
\end{equation}
We send in test particles to this extremal black hole with charge $q=Q\epsilon$, where $\epsilon \ll 1$. 
As we have pointed out in the previous section there exists a lower limit for the energy of the test particle to ensure that it is absorbed by the black hole
\begin{equation}
E\geq E_{\rm{min}}=\frac{qQ}{r_+}=\frac{\epsilon Q^2}{r_+}=\frac{M\epsilon}{8}, \label{condi1ex}
\end{equation}
where we have substituted  $q=Q\epsilon$, and used (\ref{rplus})  to express $E_{\rm{min}}$ in terms of $M$.
If the event horizon is destroyed  after the interaction, the final parameters of the spacetime satisfy
\begin{equation}
\delta_{\rm{fin}}=(M+E)^2-64(Q+q)^2\Lambda<0. \label{deltafinex}
\end{equation}
At the final stage, the mass and charge parameters of the black hole have been perturbed by the energy and the charge of the test particle. Now, we re-write (\ref{deltafinex}) by substituting $q=Q\epsilon$, and imposing the extremality condition (\ref{deltainex}).
\begin{equation}
\delta_{\rm{fin}}=E^2+2ME-M^2(\epsilon^2 + 2\epsilon)<0, \label{deltafinex1}
\end{equation}
$\delta_{\rm{fin}}$ has the form of a quadratic equation for the energy $E$, with two roots ($E_1,E_2$). $\delta_{\rm{fin}}$ will be negative if the energy of the test particle is chosen in the range  $(E_1,E_2)$ and positive outside this range. The roots of the quadratic equation are
\begin{eqnarray}
&&E_1= M\epsilon, \nonumber \\
&& E_2=-M(2+\epsilon),
\label{rootex1}
\end{eqnarray}
$\delta_{\rm{fin}}$ will identically vanish if $E=E_1$ or $E=E_2$, and will be negative for $E_2<E<E_1$. Actually we are interested in the range $0<E<E_1$ since we should assign a positive value for the energy of the test particle that we send in from infinity. A negative value is incompatible with our definition of $E$. On the other hand,  a negative value for $\delta_{\rm{fin}}$ indicates that the final parameters of the spacetime represent a naked singularity.  However we should also note that the particles with energy $E<E_{\rm{min}}$ will not be absorbed by the black hole. Therefore, if a test particle with charge $q=Q\epsilon$, has an energy in the range $E_{\rm{min}}<E<E_1$ i.e.
\begin{equation}
\frac{M \epsilon}{8} <E<M\epsilon, \label{condi2ex}
\end{equation}
it will be absorbed by the extremal black hole, and it will overcharge the extremal black hole into a naked singularity. This constitutes one of the quite rare examples in $(2+1)$ or $(3+1)$ dimensions where an extremal black hole can destroyed by test particles.

\subsection{Overcharging Nearly Extremal Black Holes}

\begin{center}
\begin{figure}
\includegraphics[scale=0.4]{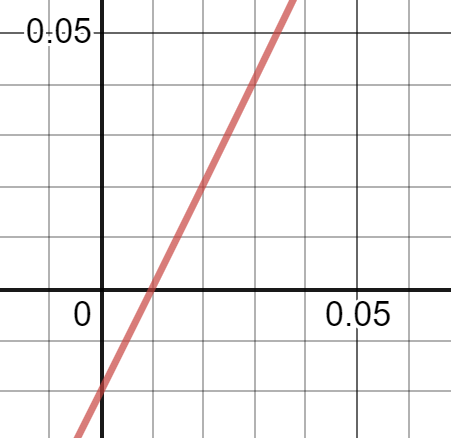}
\caption{$\delta_{\rm{fin}}$ as a function of $E$ for $M=1$, $\epsilon=\epsilon_1=0.01$.}
\label{figure}
\end{figure}
\end{center}

Another intriguing problem is to evaluate the possibility to overcharge a nearly extremal charged dilaton black hole in 2+1 dimensions. A nearly extremal black hole can be parametrised as
\begin{equation}
\delta_{\rm{in}}=M^2-64Q^2\Lambda=M^2\epsilon_1^2, \label{param}
\end{equation}
where $\epsilon_1 \ll 1$ determines the black hole's closeness to extremality. Note that the spatial location of the event horizon for this nearly extremal black hole is
\begin{equation}
r_+=\frac{M(1+\epsilon_1)}{8\Lambda}. \label{rplusnex}
\end{equation}
If a test particle has  charge $q=Q\epsilon$, the lower limit for the energy of the test particle which ensures that it is absorbed by this nearly extremal black hole is 
\begin{equation}
E_{\rm{min}}=\frac{qQ}{r_+}=\frac{\epsilon Q^2}{r_+}=\frac{\epsilon  M^2 (1-\epsilon_1^2)}{64 \Lambda r_+}=\frac{M\epsilon(1-\epsilon_1)}{8},\label{condi1nex}
\end{equation}
where we have substituted $Q^2=[M^2(1-\epsilon_1^2)]/(64\Lambda)$ using (\ref{param}). Notice that $E_{\rm{min}}$ reduces to its corresponding value derived for extremal black holes in the previous section for $\epsilon_1=0$. As we have mentioned in the previous section, if 
$\delta_{\rm{fin}}<0$ at the end of the interaction, the nearly extremal black hole will be overcharged i.e. 
\begin{eqnarray}
\delta_{\rm{fin}}&=&(M+E)^2-64(Q+q)^2\Lambda, \nonumber \\  
&=&(M+E)^2-64Q^2(1+\epsilon)^2\Lambda    <0, \label{deltafinnex0}
\end{eqnarray}
\label{deltafinnex}
where we have used  $q=Q\epsilon$. Substituting $64Q^2\Lambda=M^2(1-\epsilon_1^2)$,  the expression (\ref{deltafinnex0}) takes the form
\begin{equation}
\delta_{\rm{fin}}=M^2 +E^2+ 2ME -M^2(1-\epsilon_1^2)(1+\epsilon)^2 <0.
\end{equation}
As in the case of extremal black holes, $\delta_{\rm{fin}}$ has the form of a quadratic equation for the energy $E$, with two roots ($E_1,E_2$). $\delta_{\rm{fin}}$ will be negative if the energy of the test particle is chosen in the range  $(E_1,E_2)$ and positive outside this range . The roots of the quadratic equation are
\begin{eqnarray}
E_{1,2}&=&\frac{1}{2} \left( -2M \pm \sqrt{4M^2+4M^2(2\epsilon + \epsilon^2 -\epsilon_1^2)}\right), \nonumber \\
&=& -M \pm \sqrt{(1+\epsilon)^2 - \epsilon_1^2}. \label{rootnex1}
\end{eqnarray}
We are interested in the positive root which determines the maximum energy for a test particle with charge $q=Q\epsilon$, to overcharge the nearly extremal black hole. 
\begin{equation}
E<E_{\rm{max}}=M\left( \sqrt{(1+\epsilon)^2 - \epsilon_1^2}-1 \right). \label{condi2nex}
\end{equation}
Notice that $E_{\rm{max}}=M\epsilon$ for $\epsilon_1=0$, which is the value derived for extremal black holes. The test particle should  also satisfy the  condition (\ref{condi1nex}) so that it is absorbed by the nearly extremal black hole. Thus, a test particle satisfying
\begin{eqnarray}
&& q=\epsilon Q \nonumber \\
&&\frac{[M\epsilon(1-\epsilon_1)]}{8} < E < M\left( \sqrt{(1+\epsilon)^2 - \epsilon_1^2}-1 \right),
\end{eqnarray}
will be absorbed by a nearly extremal black hole to overcharge it into a naked singularity. 
For a numerical example, let us consider a nearly extremal black hole with $M=1$ and choose $\epsilon=\epsilon_1=0.01$. (\ref{param}) implies that $64Q^2\Lambda=0.9999$. $E_{\rm{min}}$ and $E_{\rm{max}}$ can be calculated as
\begin{eqnarray}
&&E_{\rm{min}}=\frac{M(0.01)(0.99)}{8}=0.0012375 \nonumber \\
&& E_{\rm{max}}=1(\sqrt{1.02}-1)=0.0099505
\end{eqnarray}
Let us choose $E=0.004$ and $q=\epsilon Q$ for a test particle. This particle will be absorbed by the black hole since $E>E_{\rm{min}}$. At the end of the interaction the parameters of the black hole will satisfy
\begin{eqnarray}
\delta_{\rm{fin}}&=&(M+E)^2-64Q^2\Lambda (1+\epsilon)^2 \nonumber \\
&=& (1+0.004)^2-0.9999(1+0.01)^2 \nonumber \\
&=&-0.011982 \label{numeric}
\end{eqnarray}
The negative sign for $\delta_{\rm{fin}}$ indicates that the black hole is overcharged into a naked singularity.  In figure (\ref{figure}) we have plotted $\delta_{\rm{fin}}$ as a function of $E$ for $M=1$, $\epsilon=\epsilon_1=0.01$ so that $64Q^2\Lambda=0.9999$. Though the function is quadratic it appears to be linear as we focus on a small range.

In this work we have neglected the backreaction effects. The backreaction effects will bring second order corrections to $\delta_{\rm{fin}}$. However the second order corrections cannot compensate for the destruction of the horizon for this case, since $\delta_{\rm{fin}} \sim -M\epsilon$. If we choose $E$ closer to $E_{\rm{min}}$, the absolute value of $\delta_{\rm{fin}}$ will be even larger. Therefore the violation of wccc in the interaction of charged dilaton black holes with test particles is quite generic.

\section{Violation of the third law of black hole dynamics}

Bardeen, Carter, and Hawking developed a connection between the thermodynamics and gravity by proposing the four laws of thermodynamics for stationary and axi-symmetric black holes in general relativity~\cite{bardeen}. This is achieved by setting up an analogy between the area of the black hole's event horizon  and the entropy, while  the  surface gravity $\kappa$ is identified with the temperature. Hawking proved that the area of the event horizons cannot be decreased, provided that no naked singularities exist in the outer region~\cite{area}. Thus, if the cosmic censorship conjecture is correct the second law of thermodynamics is valid for black holes. The third law, which was later proved by Israel,  states that a non-extremal black hole cannot become extremal at a finite advanced time in any continuous process~\cite{israel1}. Dadhich and Karayan  showed that the range of the allowed energy and angular momentum ratios to overspin a black hole, pinches off as one approaches extremality, which justifies the validity of the third law~\cite{dadhich}. Later, Hubeny found that one can overcharge nearly extremal black holes beyond extremality, neglecting backreaction effects~\cite{hu}. Jacobson and Sotiriou applied a similar procedure to overspin nearly extremal Kerr black holes with test particles~\cite{jacob}, which was generalised to test fields by D\"{u}zta\c{s} and Semiz~\cite{overspin}. In all these works the authors confirm that extremal black holes cannot be overcharged/overspun, while nearly-extremal black holes can be pushed beyond extremality by a discrete jump. If one attempts to approach extremality in a continuous manner, the range of the allowed energies to overcharge/overspin black holes vanishes. In this sense the overspinning/overcharging of nearly extremal black holes derived in these works, does not invalidate the third law of black hole dynamics, as it first appears to do.

The same arguments apply to the 2+1 dimensional asymptotically anti-de Sitter case.  Rocha and Cardoso proved that extremal BTZ black holes cannot be overspun by test particles~\cite{vitor}. Later, D\"{u}zta\c{s} analysed the interaction of BTZ black holes with test particles and fields to conclude  that overspinning is possible in both cases if one starts with a nearly extremal BTZ black hole. However, the range of the energies for a test particle or field to overspin a BTZ black hole vanishes as one approaches extremality; i.e. extremal black holes cannot be overspun by test particles or fields~\cite{btz}.

In this work we derived that the minimum energy $E_{\rm{min}}$ that allows the absorption of a test body by a nearly extremal charged dilaton black hole is $E_{\rm{min}}=M\epsilon(1-\epsilon_1)/(8)$. If this black hole absorbs a particle with energy $E_{\rm{max}}=M ( \sqrt{(1+\epsilon)^2 - \epsilon_1^2}-1 )$ and charge $q=\epsilon Q$, $\delta_{\rm{fin}}$ will be zero at the end of the interaction and the black hole will become extremal, which also implies that  the Hawking temperature (\ref{hawt}) will vanish. For the case of nearly extremal BTZ black holes  the range $(E_{\rm{min}},E_{\rm{max}})$ vanishes as the black holes become arbitrarily close to extremality, i.e. extremal BTZ black holes cannot be overspun~\cite{btz}. The same argument applies to  Reissner-Nordstr\"{o}m black holes studied by Hubeny~\cite{hu}, and Kerr black holes studied by Jacobson-Sotiriou~\cite{jacob} and D\"{u}zta\c{s}-Semiz~\cite{overspin}. However for the nearly extremal charged dilaton black holes which approach extremality as $\epsilon_1 \to 0$, one derives
\begin{equation}
\lim_{\epsilon_1 \to 0} \frac{E_{\rm{max}}}{E_{\rm{min}}}=\frac{8(\sqrt{(1+\epsilon)^2}-1)}{\epsilon}=8, \label{limit}
\end{equation}
which is equal to $(E_{\rm{max}}/E_{\rm{min}})$ for extremal black holes. It was already derived in the previous section that $E_{\rm{max}}$ and $E_{\rm{min}}$ reduce to the corresponding values for the extremal case for $\epsilon_1=0$. However the limit derived in (\ref{limit}) implies something further. For arbitrarily small values of $\epsilon_1$, there exists a corresponding value for the energy  $E_{\rm{max}}$ which will drive the black hole to extremality. The process can be analytically continued to drive the black hole beyond extremality. Therefore the nearly extremal charged dilaton black holes can be continuously driven to extremality and beyond, in an apparent violation of the third law of black hole thermodynamics and the wccc. The analogous limit would be equal to 1 for BTZ black holes, i.e. the corresponding range $(E_{\rm{max}},E_{\rm{min}})$ vanishes as one approaches extremality (see \cite{btz}). This implies that BTZ black holes cannot be continuously driven to extremality, unlike the charged dilaton black holes.

\section{Thermal stability}
\label{thermalstab}

The derivation that the charged dilaton black holes can be turned into naked singularities leads one to question the thermal stability of the corresponding configurations. If there exists a  thermally unstable state in the black hole-naked singularity transition, the space-time may undergo a phase transition to restore the event horizon. Thermal stability is characterised by the positivity of the heat capacity as follows \cite{J4,J5}
\begin{equation}
C_Q=T\left(\frac{\partial S}{\partial T}\right)_Q=\frac{T}{(\partial^2M/\partial T^2)_Q},
\end{equation}
where $S$ is the entropy of the black hole, and the subscript $Q$ refers to the fact the equation is evaluated at constant charge. It turns out that there are two types of instabilities that lead to phase transitions \cite{sheykhi}. If there exists a point where the heat capacity vanishes, the spacetime undergoes type-one phase transitions. Type-two phase transitions take place if there exists a point where the heat capacity diverges. Previously, the possibility of the occurrence of phase transitions has been analysed for dilatonic BTZ black holes \cite{dehghani}. In this section we carry out a similar analysis for charged dilatonic black holes in $(2+1)$ dimensions. To ensure that the black hole is stable, we demand that
\begin{equation}
\left(\frac{\partial S}{\partial T}\right)_Q >0.
\end{equation}
Using Hawking-Bekenstein relation, one derives that
\begin{equation}
S=\frac{A}{4}=2\pi r_{+},
\end{equation}
Notice that the Hawking temperature given in (\ref{hawt}) can also be written in the form
\begin{equation}
T_{\rm{H}}=\frac{1}{4\pi} \left(8 \Lambda -\frac{M}{r_+} \right),
\end{equation}
which implies that 
\begin{equation}
\frac{\partial T}{\partial r_+}=\frac{1}{4\pi} \left(\frac{M}{r_+^2} \right).
\end{equation}
This leads to
\begin{equation}
\left. \left(\frac{\partial S}{\partial T}\right)_Q =\left(\frac{\partial S}{\partial r_+}\right)\left(\frac{\partial r_+}{\partial T} \right)\right\vert_Q =\frac{2 \pi^2 r_+^2}{M}.
\label{stable}
\end{equation}
The expression (\ref{stable}) implies that the heat capacity does not vanish unless $r_+$ vanishes itself, and there is no point of divergence. Therefore we conclude that the heat capacity is positive definite and type-one or type-two phase transitions will not occur.

\section{Summary and conclusions}

In this paper, we applied a  test of the weak cosmic censorship conjecture for extremal and nearly-extremal 2+1 dimensional charged dilaton black holes, as they interact with test charged particles. After performing our analysis, we found that  the  final parameters of the extremal black hole after particle absorption  indicate the formation of a naked singularity, provided that the energy of the particle  is in the range $(M \epsilon)/(8) <E<M\epsilon$. Previously, similar analysis were applied for BTZ black holes, which indicated that extremal BTZ black holes cannot be overspun by test particles~\cite{vitor} or fields~\cite{btz}. We also attempted to overcharge nearly-extremal dilaton  black holes, and derived that the formation of naked singularities is possible analogous to the case of nearly-extremal BTZ black holes. Throughout our analysis, we ignored the backreaction effects which would bring second order corrections to $\delta_{\rm{fin}}$.  However, the numerical value of $\delta_{\rm{fin}}$ in (\ref{numeric}) indicates that the second order corrections cannot compensate for the destruction of the horizon, since $\delta_{\rm{fin}}\sim -M\epsilon$. Therefore the overcharging of dilaton black holes derived in this work, is quite generic.

We argued that for every charged dilaton black hole arbitrarily close to extremality, there exists a range of energies for a test particle which ensures that it is absorbed by the black hole to drive it to extremality and beyond. Therefore charged dilaton black holes can be continuously driven to extremality unlike BTZ, Kerr and Reissner-Nordstr\"{o}m black holes, for which the range of allowed energies vanishes as one approaches extremality. Overspinning and overcharging of nearly extremal BTZ, Kerr and  Reissner-Nordstr\"{o}m black holes occur by a discrete jump, while it turns out to be a continuous process for charged dilaton black holes in 2+1 dimensions. Thus, the third law of black hole dynamics is also violated in the interaction of charged dilaton black holes with test particles. 

The fact that the formation of naked singularities appears generic, leads one to question the thermal stability of the intermediate states since a possible phase transition can occur and the formation of the naked singularities can be prevented. For that purpose, we searched for points at which the heat capacity vanishes or diverges, in section (\ref{thermalstab}). Our analysis implies that the heat capacity does not vanish or diverge; i.e. phase transitions of type-one or type-two will not occur. The charged dilaton black holes are thermally stable in the relevant range 
\[
T_{\rm{H}} \geq 0 \Leftrightarrow M \geq 8 Q\sqrt{\Lambda} \Leftrightarrow r_+ \geq M/8\Lambda
\]
where one can define thermodynamics. Therefore the formation of naked singularities cannot be prevented by phase transitions.

In this work we have perturbed charged dilaton black holes with test particles. The advantage of using test particles lies in the simplicity of geodesic equations. One can wander whether we should expect similar results for test fields. For test fields, one has to solve Klein-Gordon, Dirac, or Maxwell equations in the relevant background. There exists an upper bound for the energy of the incident wave to destroy the event horizon, as in the case of test particles. However, we cannot be sure if there exists a lower bound for energy which ensures that the test field to is absorbed by the black hole.  The lower bound will exist if superradiance occurs for the incident test field. Naively, we expect superradiance to occur for bosonic fields and to be absent for fermionic fields. If this is the case, we could expect results similar to test particles for bosonic test fields and a more generic destruction of the event horizon for fermionic fields. This is due to the fact that since $E_{\rm{min}}=0$ if superradiance does not occur. However, none of these can be taken for granted without an explicit solution for test fields in the background geometry of the black hole.

\end{document}